\documentclass[prb,aps,nobalancelastpage,showpacs,amssymb,twocolumn,citeautoscript]{revtex4}    %PRB(R)
\usepackage{mathrsfs}
\usepackage[figuresright]{rotating}
\usepackage{amsmath}
\usepackage{amssymb}
\usepackage{graphicx}% Include figure files
\usepackage{color}
\usepackage{dcolumn}% Align table columns on decimal point
\usepackage{bm}% bold math
\usepackage{multirow}

\makeatletter

\newcommand{\Rmnum}[1]{\expandafter\@slowromancap\romannumeral #1@}
\makeatother

\begin{document}

\title{$\mu$SR Investigation and Suppression of $T_C$ by overdoped Li in Diluted Ferromagnetic Semiconductor Li$_{1+y}$(Zn$_{1-x}$Mn$_x$)P}

\author{Huiyuan Man$^{1}$, Xin Gong$^{1}$, Guoxiang Zhi$^{1}$, Shengli Guo$^{1}$, Cui Ding$^{1}$, Quan Wang$^{1}$, T. Goko$^{2,3}$, L. Liu$^{2}$, B. A.
Frandsen$^{2}$, Y. J. Uemura$^{2}$, H. Luetkens$^{3}$, E.
Morenzoni$^{3}$, C.Q. Jin$^{4}$, T. Munsie$^{5}$, G.M. Luke$^{5,6}$,
Hangdong Wang$^{7}$, Bin Chen$^{7}$ and F. L. Ning$^{1}$}

\email{ningfl@zju.edu.cn}

\affiliation{$^{1}$Department of Physics, Zhejiang University,
Hangzhou 310027, China} \affiliation{$^{2}$Department of Physics,
Columbia University, New York, New York 10027, USA}
\affiliation{$^{3}$Paul Scherrer Institute, Laboratory for Muon Spin
Spectroscopy, CH-5232 Villigen PSI, Switzerland}
\affiliation{$^{4}$Beijing National Laboratory for Condensed Matter
Physics, and Institute of Physics, Chinese Academy of Sciences,
Beijing 100190, China}\affiliation{$^{5}$Department of Physics and
Astronomy, McMaster University, Hamilton, Ontario L8S4M1,
Canada}\affiliation{$^{6}$Canadian Institute for Advanced Research,
Toronto, Ontario M5G1Z8, Canada}
 \affiliation{$^{7}$Department of Physics,
Hangzhou Normal University, Hangzhou 310016, China}

\date{\today}% It is always \today, today,


\begin{abstract}
We use muon spin relaxation ($\mu$SR) to investigate the magnetic
properties of a bulk form diluted ferromagnetic semiconductor (DFS)
Li$_{1.15}$(Zn$_{0.9}$Mn$_{0.1})$P with $T_C$ $\sim$ 22 K. $\mu$SR
results confirm the gradual development of ferromagnetic ordering
below $T_C$ with a nearly 100\% magnetic ordered volume. Despite its
low carrier density, the relation between static internal field and
Curie temperature observed for Li(Zn,Mn)P is consistent with the
trend found in (Ga,Mn)As and other bulk DFSs, indicating these
systems share a common mechanism for the ferromagnetic exchange
interaction. Li$_{1+y}$(Zn$_{1-x}$Mn$_x$)P has the advantage of
decoupled carrier and spin doping, where Mn$^{2+}$ substitution for
Zn$^{2+}$ introduces spins and Li$^+$ off-stoichiometry provides
carriers. This advantage enables us to investigate the influence of
overdoped Li on the ferromagnetic ordered state. Overdoping Li
suppresses both $T_C$ and saturation moments for a certain amount of
spins, which indicates that more carriers are detrimental to the
ferromagnetic exchange interaction, and that a delicate balance
between charge and spin densities is required to achieve highest
$T_C$.
\end{abstract}


\maketitle

\section{Introduction}

The successful fabrication of III-V (In,Mn)As and (Ga,Mn)As through
low temperature molecular beam epitaxy (LT-MBE) has generated
extensive research into diluted ferromagnetic semiconductors (DFS)
\cite{InMnAs,GaMnAs,Chambers,Dietl1,Jungwirth,Samarth,Zutic,Dietl2}.
Theoretically, it has been proposed that $T_{C}$ can reach room
temperature with sufficient spin and carrier
density\cite{ZenerModel}. After almost two decades' efforts, the
Curie temperature $T_{C}$ of (Ga,Mn)As thin films has been improved
to as high as 200 K\cite{GaMnAs_190K,ZhaoJH1,ZhaoJH2}. One of the
major intrinsic difficulties is the low solid solubility of
$\mbox{Mn}^{2+}$ substitution for $\mbox{Ga}^{3+}$, which makes it
difficult to enhance the concentration of Mn while controlling the
homogeneity of thin films. Mn$^{2+}$ substitution for Ga$^{3+}$
provides not only local moments but also hole carriers. During the
fabrication of thin films, some Mn$^{2+}$ easily get into the
interstitial sites and become a double donor, which makes it
difficult to determine precisely the amount of Mn that substitutes
for Ga at the ionic sites \cite{Jungwirth}. Investigating some DFS
systems with more controllable spin and carrier density might be
helpful in understanding the general mechanism of ferromagnetism.

Recently, several bulk form DFS families that are derivatives of
FeAs-based high temperature superconductors have been reported
\cite{(BaK)(CdMN)2As2_YXJ,(BaK)(ZnMn)2As2_JCQ,(LaBa)(ZnMn)AsO_DC,
(LaCa)(ZnMn)SbO_JCQ,(LaSr)(CuMn)SO_YXJ,32522_MHY,Li(ZnMn)As_JCQ,
Li(ZnMn)P_JCQ,Li(ZnCr)As_WQ,(LaSr)(ZnFe)AsO_LJC}. Among them, the
$T_C$ of 122 type bulk form (Ba,K)(Zn,Mn)$_2$As$_2$ reaches $\sim$
180K \cite{(BaK)(ZnMn)2As2_JCQ}, which is already comparable with
the record $T_C$ of (GaMn)As \cite{ZhaoJH1,ZhaoJH2}. $\mu$SR
measurements have shown that the relation between static internal
field and Curie temperature observed in I-II-V Li(Zn,Mn)As
\cite{Li(ZnMn)As_JCQ}, 1111 type (La,Ba)(Zn,Mn)AsO
\cite{(LaBa)(ZnMn)AsO_DC}and 122 type (Ba,K)(Zn,Mn)$_2$As$_2$
\cite{(BaK)(ZnMn)2As2_JCQ} DFSs all fall into the scaling observed
in (Ga,Mn)As thin films\cite{Dunsiger}, indicating they all share a
common mechanism for the ferromagnetic exchange interaction. The
availability of a specimen in bulk form also enables the NMR
measurements of DFS. Through $^7$Li NMR of Li(Zn,Mn)P
\cite{Li(ZnMn)P_JCQ}, Ding \emph{et al.} successfully identified
Li(Mn) sites that have Mn$^{2+}$ at N.N. (nearest neighbor) sites,
and found that the spin lattice relaxation rate $\frac{1}{T_{1}}$ is
temperature independent above $T_C$, i.e., $\frac{1}{T_{1}}$ $\sim$
400 s$^{-1}$, indicating that Mn spin-spin interaction extends over
many unit cells with an interaction energy $\mid$J$\mid$ $\sim$ 100
K \cite{LiZnMnP_NMR}. This explains why DFS exhibits a relatively
high $T_C$ with such a low density of Mn.

\begin{figure}[h]
\includegraphics[width=8cm]{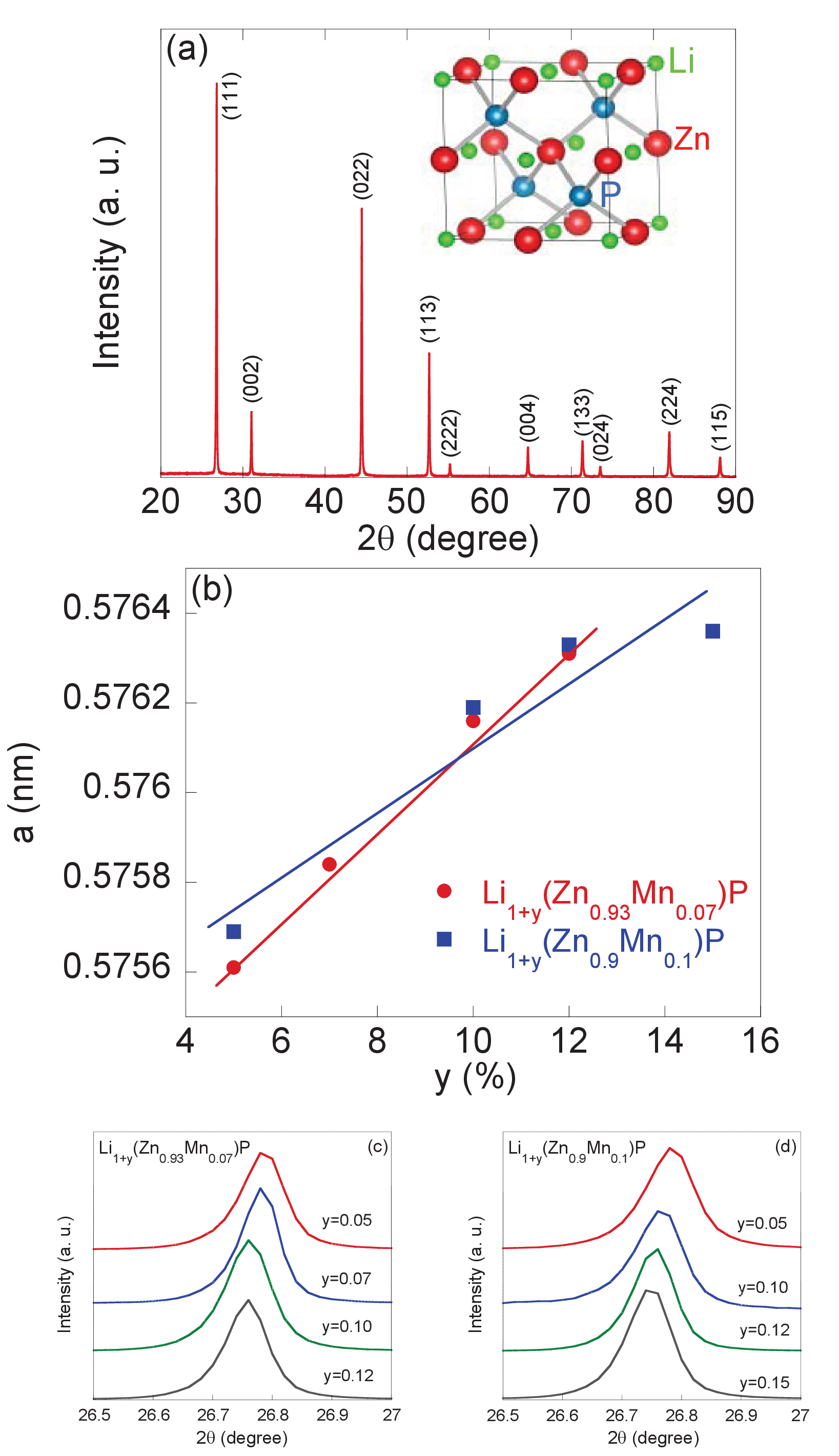}
\caption{(Color online) (a) Powder x-ray diffraction patterns of the
parent compound LiZnP with the Miller indices. The inset shows the
cubic lattice of LiZnP. (b) The lattice constants of
Li$_{1+y}$(Zn$_{0.93}$Mn$_{0.07}$)P and
Li$_{1+y}$(Zn$_{0.9}$Mn$_{0.1}$)P, solid lines are guides for the
eye. (c) and (d) Amplified plot of (111) peak for
Li$_{1+y}$(Zn$_{0.93}$Mn$_{0.07}$)P and
Li$_{1+y}$(Zn$_{0.9}$Mn$_{0.1}$)P.}
\end{figure}
The other advantage of these new bulk DFSs is the decoupling of
carrier and spin doping. (La,Ba)(Zn,Mn)AsO DFS with $T_C$ $\sim$ 40
K \cite{(LaBa)(ZnMn)AsO_DC}, for example, is achieved by doping
Mn$^{2+}$ and Ba$^{2+}$ into the direct gap semiconductor LaZnAsO,
where Mn$^{2+}$ substitution for Zn$^{2+}$ and Ba$^{2+}$
substitution for La$^{3+}$ introduce spin and hole carriers,
respectively. Both Mn and Ba are chemically stable elements, and the
concentrations of charge and spin can be precisely controlled. Very
recently, by controlling the Li concentration and by doping Mn into
the direct gap semiconductor LiZnP (gap $\sim2.1$
eV\cite{LiZnPgap,LiZnPgap2}), Li(Zn,Mn)P \cite{Li(ZnMn)P_JCQ} has
been found to experience a transition into ferromagnetic state below
$\sim$ 34K with a low carrier density of 10$^{16}$/cm$^3$. This
carrier density is $\sim$ 3 orders smaller than that of (Ga,Mn)As,
Li(Zn,Mn)As and (Ba,K)(Zn,Mn)$_2$As$_2$, leaving the open questions
about the mechanism behind ferromagnetism of Li(Zn,Mn)P.

In this paper, we use $\mu$SR to investigate the ferromagnetism of a
I-II-V DFS Li(Zn,Mn)P with $T_C$ = 22 K. Our $\mu$SR results
demonstrate that a nearly 100\% ferromagnetic ordered volume
develops below $T_C$, indicating the homogenous distribution of
Mn$^{2+}$ atoms. This result is consistent with NMR measurements of
the same specimen. We also investigated the spin dynamics of Mn
spins, and found that Li(Zn,Mn)P also falls into the same scalings
of internal field and $T_C$ observed for (Ga,Mn)As and other bulk
DFSs, indicating that Li(Zn,Mn)P belongs to the DFS families that
share a common mechanism of ferromagnetic exchange interaction
despite its much lower carrier density. Furthermore, taking
advantage of decoupled carrier and spin doping, we studied the
carrier doping effect on the ferromagnetic ordered state. Our
experimental results show that overdoping Li suppresses both $T_C$
and the saturation moments. In other words, too many carriers are
harmful to the development of ferromagnetic ordering in a similar
way as too few.

\begin{figure}
\includegraphics[width=8cm]{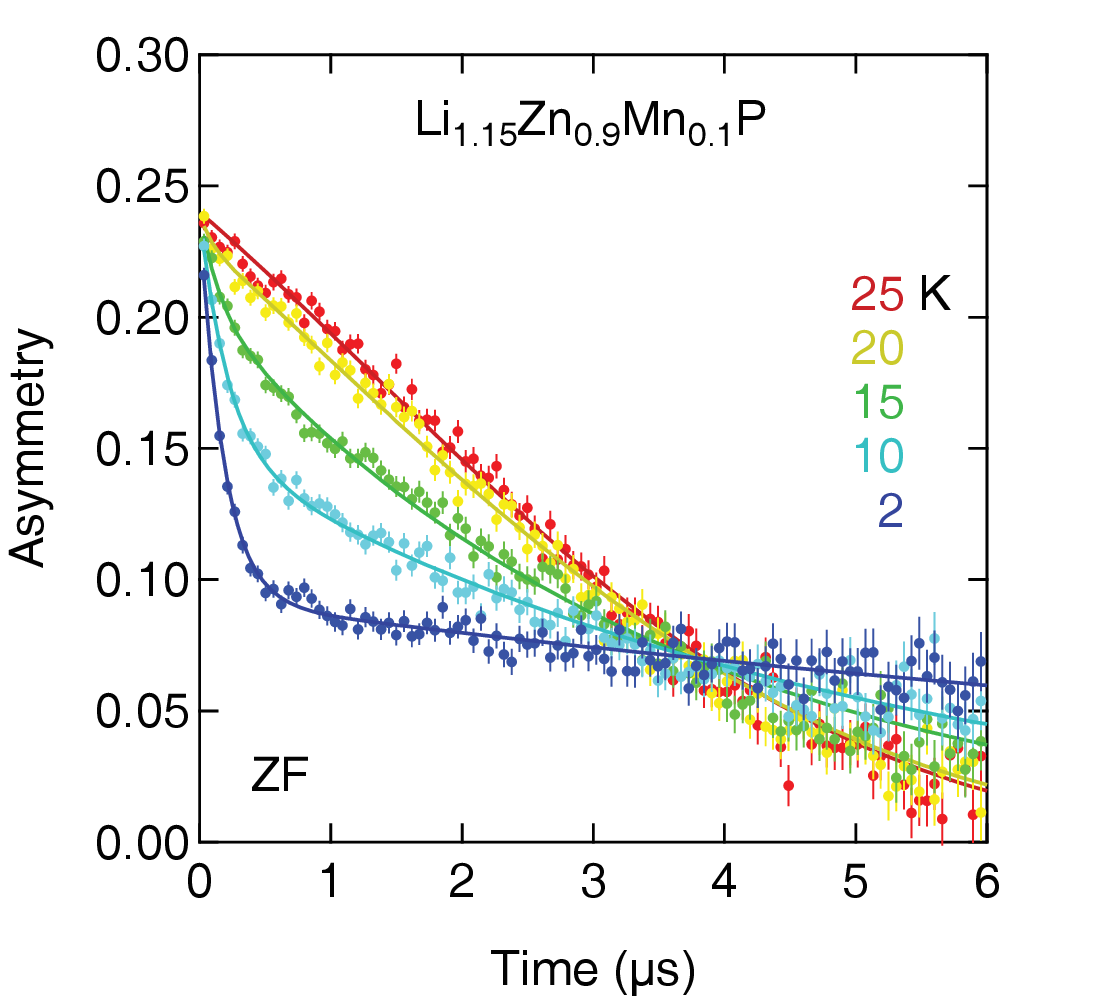}
\caption{(Color online) Time spectra of zero field $\mu\mbox{SR}$
measurements for Li$_{1.15}$(Zn$_{0.9}$Mn$_{0.1}$)P with $T_C$
$\sim$ 22K. The solid lines are fits to a two component relaxation
function for dilute spin systems \cite{Li(ZnMn)As_JCQ}.}
\end{figure}

\section{Experiments}

The Li$_{1+y}$(Zn$_{1-x}$Mn$_x$)P polycrystalline specimens were
synthesized by the solid state reaction method. High purity elements
of Li (99.9\%), Zn (99.9\%), Mn (99.99\%), and P (99\%) were mixed
and slowly heated to 450 $^{\circ}\mbox{C}$ in evacuated silica
tubes, and held at 450 $^{\circ}\mbox{C}$ for 48 hours. After
cooling down to room temperature, the mixture was ground thoroughly,
then pressed into pellets and heated again to ensure the complete
reaction. The handling of materials was performed in a high-purity
argon filled glove box (the percentage of $\mbox{O}_{2}$ and
$\mbox{H}_{2}\mbox{O}$ was $\leq$ 0.1 ppm) to protect it from air.
Powder X-ray diffraction was performed at room temperature using a
PANalytical X-ray diffractometer (Model EMPYREAN) with monochromatic
$\mbox{CuK}_{\alpha1}$ radiation. The dc magnetization measurements
were conducted on Quantum Design SQUID (superconducting quantum
interference device). Zero-field (ZF) and weak-transverse-field
(WTF) muon spin relaxation measurements were performed at PSI and
TRIUMF. The specimen used for $\mu$SR study in this work is the same
piece as used in the NMR study \cite{LiZnMnP_NMR}.

\section{Results and discussion}

\begin{figure}
\includegraphics[width=8cm]{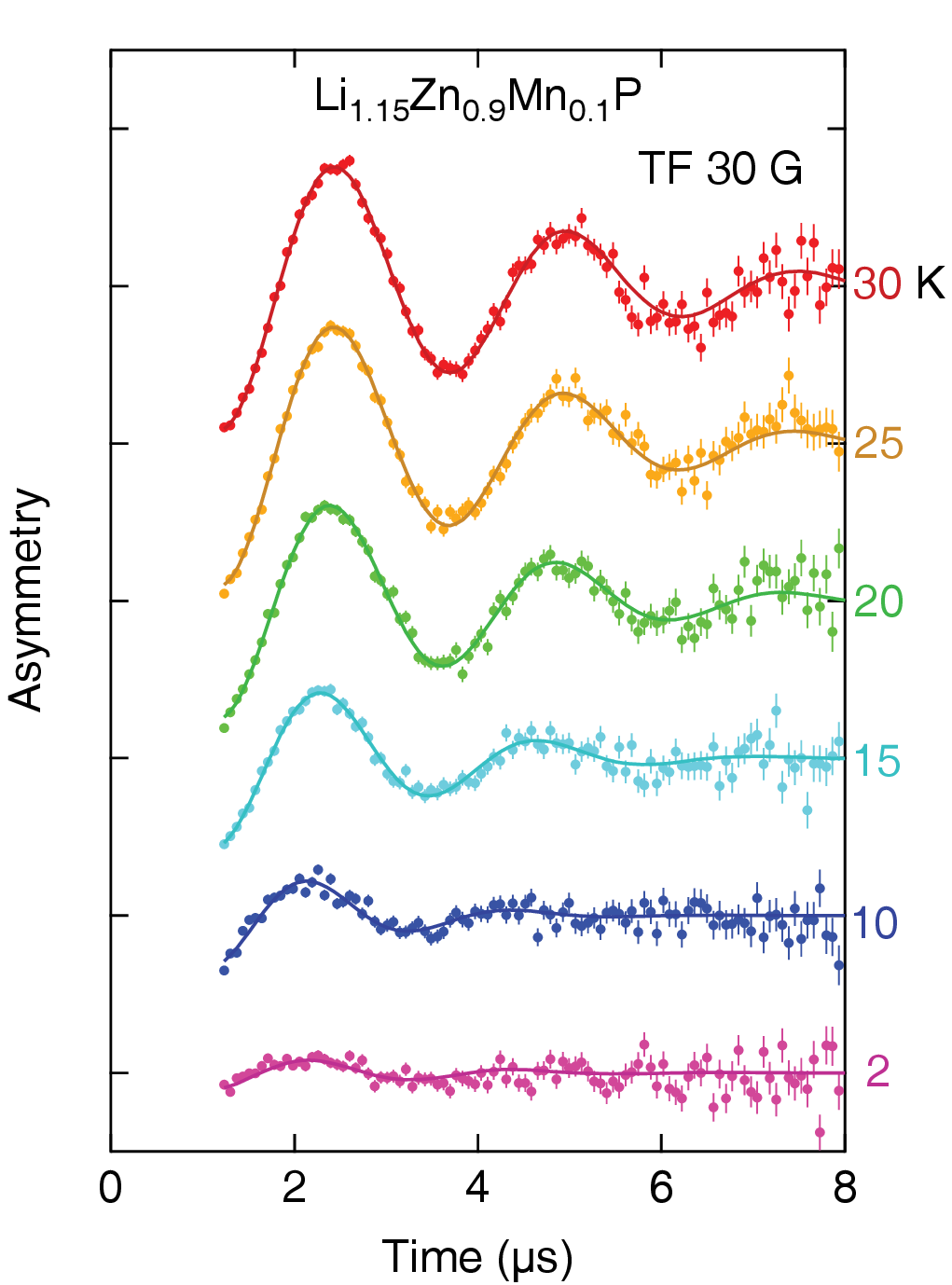}
\caption{(Color online) $\mu$SR time spectra of
Li$_{1.15}$(Zn$_{0.9}$Mn$_{0.1}$)P in a weak-transverse-field of 30
Oe. The oscillation amplitude corresponds to the paramagnetic volume
faction.}
\end{figure}
In Fig 1(a), we show powder X-ray diffraction patterns of the parent
compound LiZnP with the Miller indices. Bragg peaks can be well
indexed into a cubic structure with a space group F\={4}3m,
identical to the zinc blende
GaAs\cite{Li(ZnMn)As_JCQ,Li(ZnMn)P_JCQ}. Doping excess Li and Mn
into the parent compound does not change the crystal structure and
no impurities have been observed. The lattice constants $a$ of
Li$_{1+y}$(Zn$_{0.93}$Mn$_{0.07}$)P and
Li$_{1+y}$(Zn$_{0.9}$Mn$_{0.1}$)P are shown in Fig. 1(b). The
evolution of lattice constant $a$ follows the Vegard law, indicating
the successful Li doping. This trend can also be seen from the
amplified (111) peaks shown in Fig. 1(c) and (d), which shift
systematically towards smaller $2\theta$ with increasing $y$,
suggesting that Li atoms are indeed incorporated into the lattice.
This results in the enlargement of the unit cell. In addition, it
has been shown that the NMR line width of LiZnP is only $\sim$4 KHz
\cite{LiZnMnP_NMR}. This is comparable to the line width of pure Cu
metal, indicating the high quality of these polycrystals. We do not
observe signals arising from Li atoms that enter Zn sites which
would give rise to additional NMR peaks due to the different
electrical environment they are located in \cite{LiZnMnP_NMR}.

\begin{figure}
\includegraphics[width=8cm]{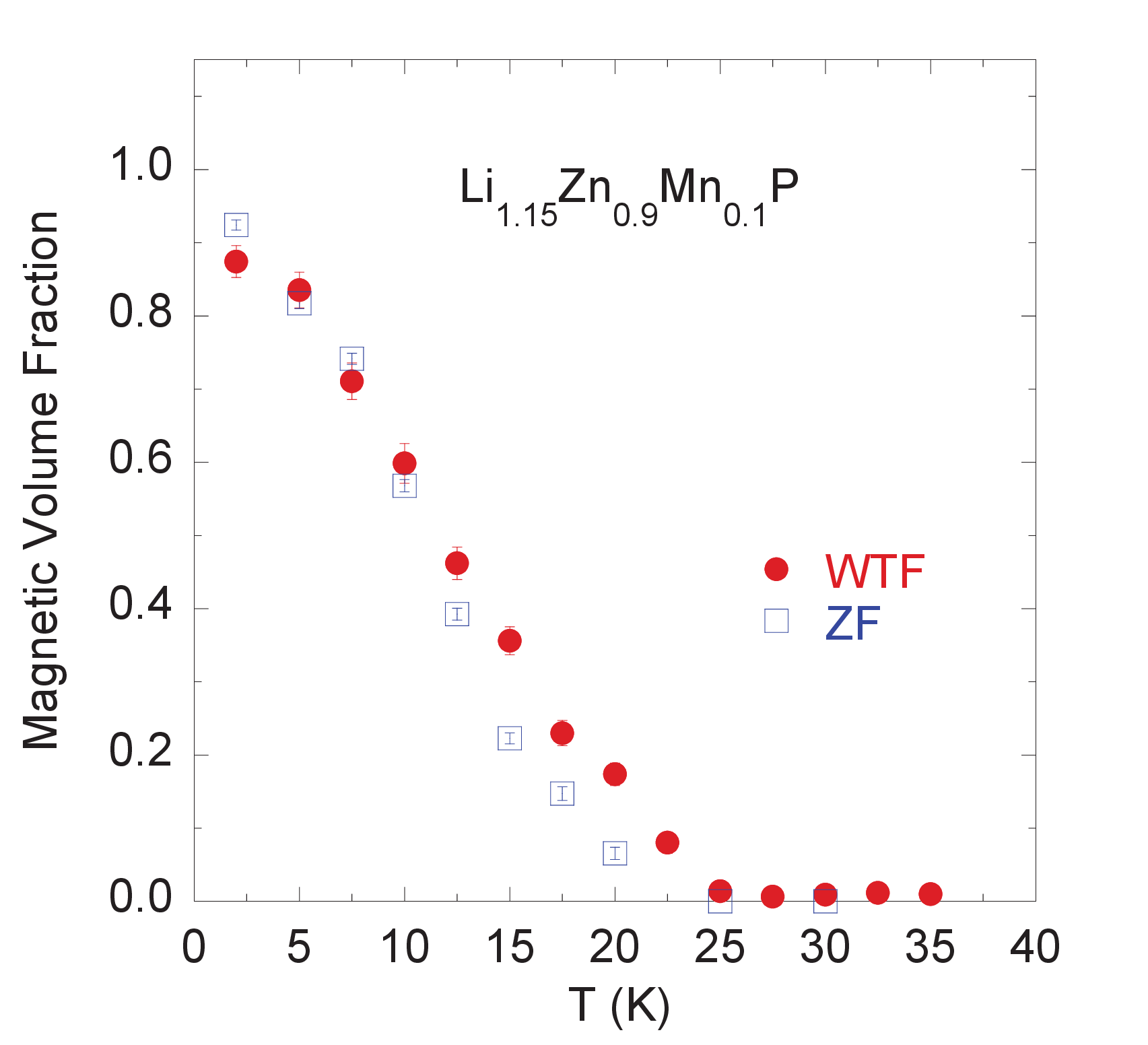}
\caption{(Color online) Temperature dependent magnetic volume
fraction derived from ZF ($\square$) and WTF-$\mu$SR ($\bullet$)
measurements for Li$_{1.15}$(Zn$_{0.9}$Mn$_{0.1}$)P.}
\end{figure}

In Fig. 2, we show the time spectra of the Zero-field (ZF) $\mu$SR
on the polycrystalline sample Li$_{1.15}$(Zn$_{0.9}$Mn$_{0.1}$)P
which has a $T_C$ = 22 K. It can be seen that the time spectra
clearly displays an increase in relaxation rate below T = 15 K. We
note that no clear oscillation is observed even at 2 K; a similar
situation is observed in (Ga,Mn)As \cite{Dunsiger} and Li(Zn,Mn)As
\cite{Li(ZnMn)As_JCQ} as well. This can be attributed to the random
distribution of Mn moments in real space because Mn substitution for
Zn is random. This makes the local field at the muon site highly
random even in the ferromagnetic ground state, and ZF precession
signals are subsequently strongly damped. The random distribution of
Mn moments has also been shown by Li(Mn) NMR lineshapes, which
displays a large distribution between 61 MHz and 66 MHz
\cite{LiZnMnP_NMR}. The broad NMR line arises from the distribution
of hyperfine fields at Li(Mn) sites, which can be as large as $\sim$
0.3 Tesla\cite{LiZnMnP_NMR}. Note that the transferred hyperfine
coupling between Li nuclear spins and Mn electrons is much weaker
than those between Zn/Mn and As. Currently, we do not know the exact
muon sites in Li(Zn,Mn)P, but relatively large hyperfine fields at
muon sites are likely. We employed a two component function to
analyze the ZF spectra (the fitting function is identical to the one
used for Li(Zn,Mn)As, as explained in the method section of ref.
13). One component is for a static magnetic field with a Lorentzian
distribution. This is expected for dilute Mn moments randomly
substituting Zn sites, representing the magnetically ordered volume.
The other component is an exponential function, representing the
volume fraction of fluctuating paramagnetism. The derived
ferromagnetic ordered volume fraction is shown in Fig. 4.

\begin{figure}
\includegraphics[width=8cm]{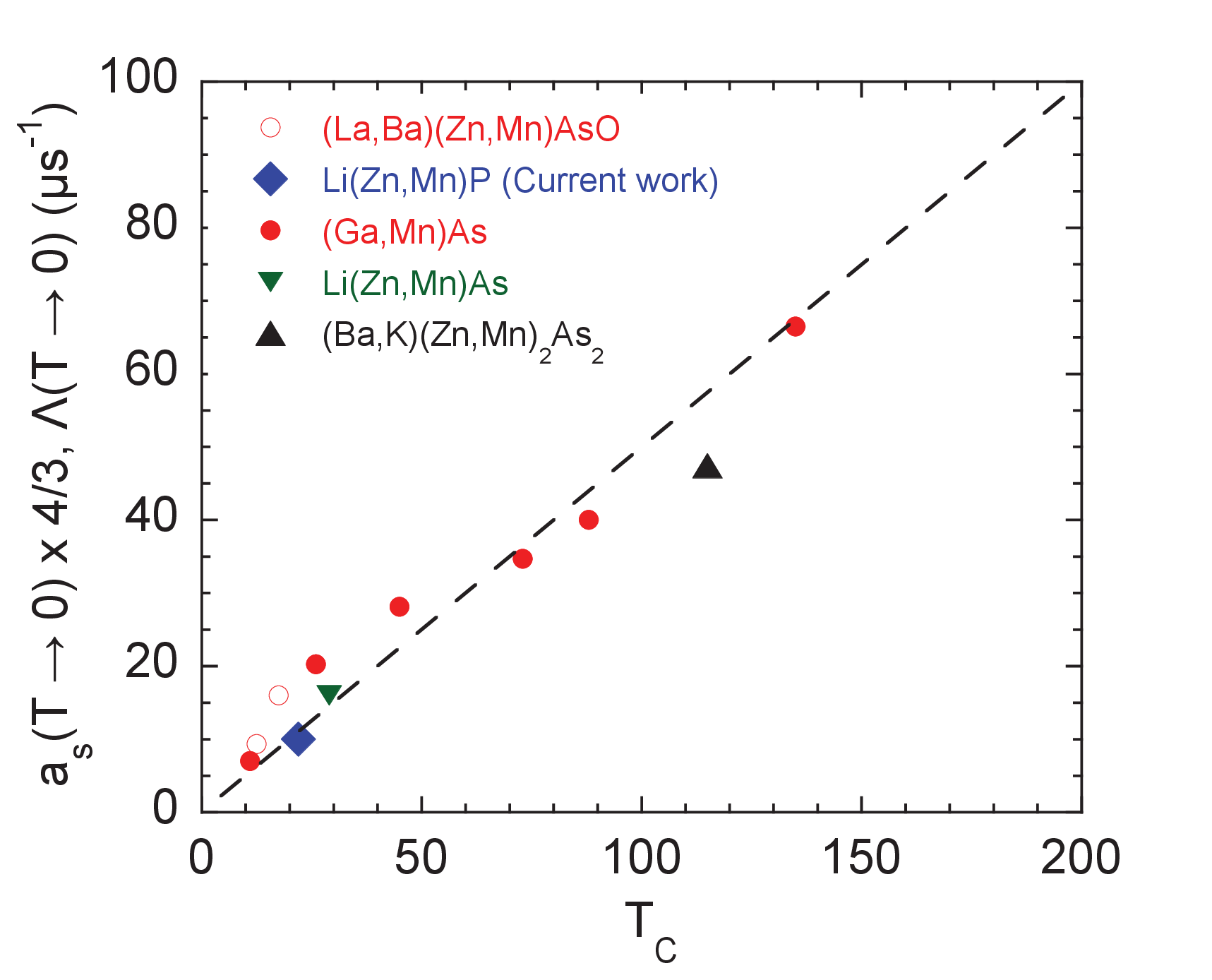}
\caption{(Color online) Correlation between the static internal
field parameter $a_s$ determined at 2 K by zero-field $\mu$SR versus
Curie temperature $T_C$ observed in (Ga,Mn)As (Ref. 23), Li(Zn,Mn)As
(Ref. 13), (La,Ba)(Zn,Mn)AsO (Ref. 16), (Ba,K)(Zn,Mn)$_2$As$_2$
(Ref. 20) and current work of Li(Zn,Mn)P. The nearly linear
correlation indicates a common mechanism for the ferromagnetic
exchange interaction; dashed line is a guide for the eye.}
\end{figure}

\begin{table*}
\normalsize
 \caption{Curie temperature ($T_C$), Weiss temperature ($\theta$), the
effective moment ($M_{eff}$), and the saturation moment ($M_{sat}$,
the value measured at $T=5$ K and H = 500 Oe) for
Li$_{1+y}$(Zn$_{1-x}$Mn$_x$)P.}
  \begin{center}
  \renewcommand{\multirowsetup}{\centering}
    \begin{tabular}{|c|c|c|c|c|c|c|} \hline
    \multirow{5}{3.5cm}{Li$_{1+y}$(Zn$_{0.93}$Mn$_{0.07}$)P}
    &{y$$ (excess Li) }&{$T_C$ (K)} &{$\theta$ (K)}&{$M_{sat}$ ($\mu_B$/Mn)}&{$M_{eff}$ ($\mu_B$/Mn)}&{Coercivity (Oe)}  \\ \hline
    &{0.07}&{25} &{24.5}&{0.61}&{3.40}&{20}  \\
    &{0.10}&{24} &{23.0}&{0.32}&{2.40}&{16}  \\
    &{0.15}&{24} &{23.0}&{0.37}&{2.02}&{25}  \\ \hline
    \multirow{4}{3.5cm}{Li$_{1+y}$(Zn$_{0.9}$Mn$_{0.1}$)P}
    &{0.07}&{22} &{19.8}&{0.19}&{1.94}&{24}  \\
    &{0.10}&{25} &{23.4}&{0.41}&{2.52}&{14}  \\
    &{0.12}&{22} &{18.4}&{0.23}&{2.34}&{20}  \\
    &{0.15}&{22} &{17.7}&{0.22}&{2.29}&{20}  \\\hline

\end{tabular}
  \end{center}
\end{table*}

In Fig. 3, we show the $\mu$SR time spectra measured in a weak
transverse field (WTF) of 30 Oe. As stated previously, the average
internal magnetic field at the muon sites should be much larger than
the applied field, and the oscillation amplitude is therefore an
indicator of the paramagnetic volume \cite{Uemura}. WTF-$\mu$SR
measurements provide direct information of the magnetic volume
fraction \cite{Uemura}. Clearly, the amplitude of oscillation
becomes smaller with decreasing $T$, indicating the suppression of
the paramagnetic volume. We show the results of size of magnetically
ordered volume derived from WTF-$\mu$SR in Fig. 4, and compare it
with those derived from the measurements in ZF. Both are in good
agreement, supporting the validity of our analysis of the ZF-$\mu$SR
spectra using a two-component function. We note that the growth of
ferromagnetic ordered volume below T$_C$ in Li(Zn,Mn)P is more
gradual than previous reported Li(Zn,Mn)As \cite{Li(ZnMn)As_JCQ} and
122 type (Ba,K)(Zn,Mn)$_2$As$_2$ \cite{(BaK)(ZnMn)2As2_JCQ}, this is
primarily due to the distribution of T$_C$ arising from sample
synthesis. We have also conducted longitudinal-field $\mu$SR at 200
Oe (not shown), and confirmed the relaxation observed in this
material is due mostly to static magnetic field. The ordered volume
fraction starts to grow below $\sim$ 22K, indicating a ferromagnetic
transition that takes place at $T_{C}\sim$22K, which is consistent
with the magnetization measurements. Magnetic volume fraction
reaches up to nearly $100\%$ at the base temperature of 2 K,
implying that static magnetic order develops in almost the entire
sample volume. This result is again consistent with the NMR
observation that no NMR signals arising from magnetic impurities or
Mn clusters are observed \cite{LiZnMnP_NMR}. We believe that further
improvement in fabrication and heating procedures will improve the
homogeneity of ferromagnetism in this material.

\begin{figure}
\includegraphics[width=8cm]{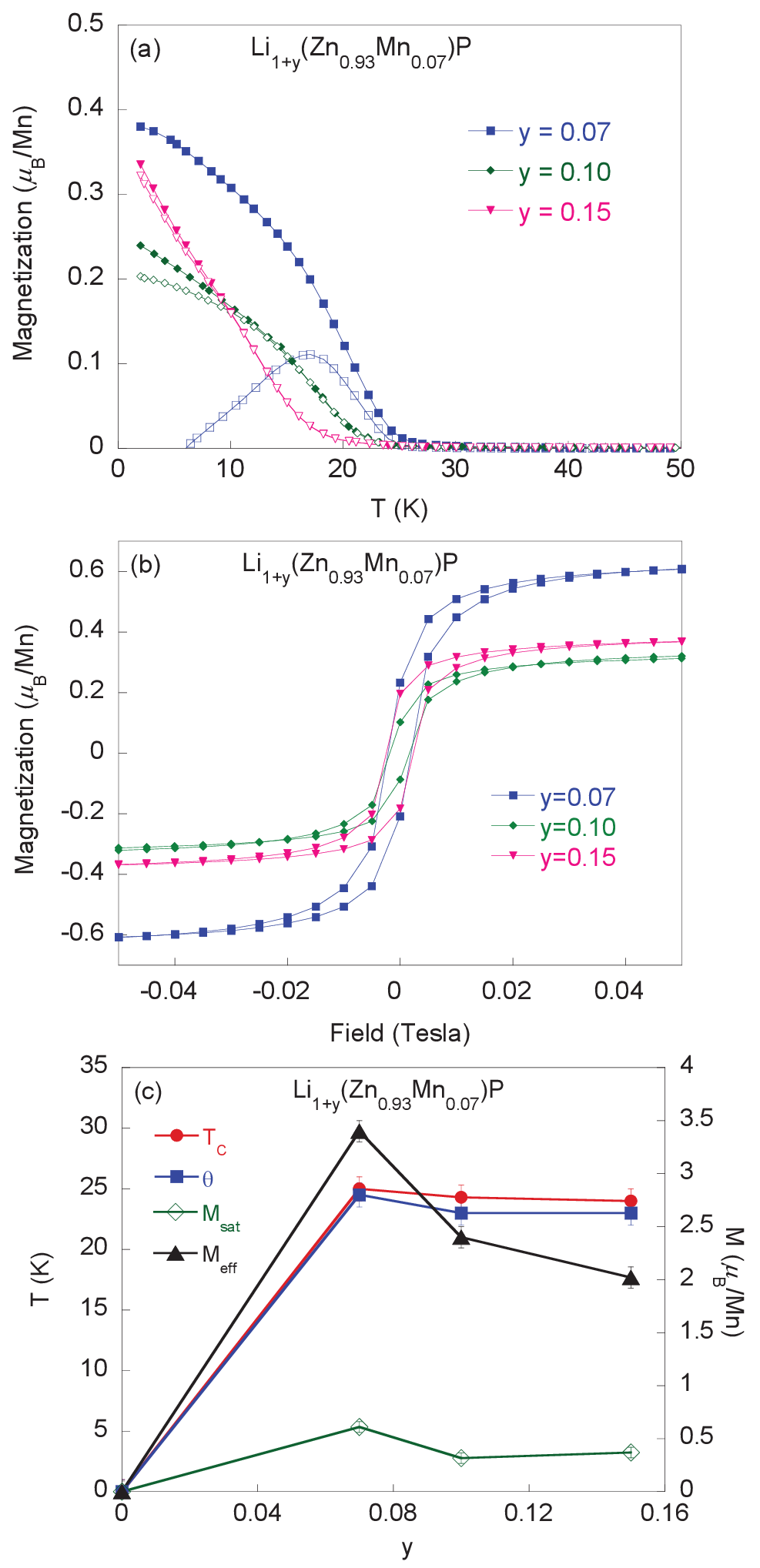}
\caption{(Color online) (a) dc magnetization for
Li$_{1+y}$(Zn$_{0.93}$Mn$_{0.07}$)P ($y=0.07$, 0.10 and 0.15) in
both zero field cooling (ZFC, open symbols) and field cooling (FC,
solid symbols) modes under $B_{ext}$ = 50 Oe. (b) Isothermal
magnetization measured at 5 K for
Li$_{1+y}$(Zn$_{0.93}$Mn$_{0.07}$)P ($y=0.07$, 0.10 and 0.15)
specimens. (c) Curie temperature $T_{C}$, Weiss temperature
$\theta$, saturation moments $M_{sat}$ and effective moments
$M_{eff}$ versus excess Li$^+$ concentrations y of
Li$_{1+y}$(Zn$_{0.93}$Mn$_{0.07}$)P.}
\end{figure}

We can extract the relaxation rate, $a_s$, from the fitting of ZF
time spectra. $a_s$ is proportional to the individual ordered moment
size multiplied by the moment concentration. Similar to the case of
Li(Zn,Mn)As \cite{Li(ZnMn)As_JCQ}, we plot $a_s$ at the lowest
measured temperature versus Curie temperature $T_C$ in Fig. 5. A
factor 4/3 is multiplied to the parameter $a_s$ for all bulk DFSs
\cite{Li(ZnMn)As_JCQ,(BaK)(ZnMn)2As2_JCQ,(LaBa)(ZnMn)AsO_DC}, to
adjust the difference from the simple exponential decay rate adopted
in (Ga,Mn)As \cite{Dunsiger}. $a_s$ falls into the linear relation
observed for (Ga,Mn)As and other DFS families. The good agreement
implies that all these DFS systems share a common mechanism for the
ferromagnetic exchange interaction, including the current Li(Zn,Mn)P
system, despite its much lower carrier density \cite{Li(ZnMn)P_JCQ}.

\begin{figure}
\includegraphics[width=8cm]{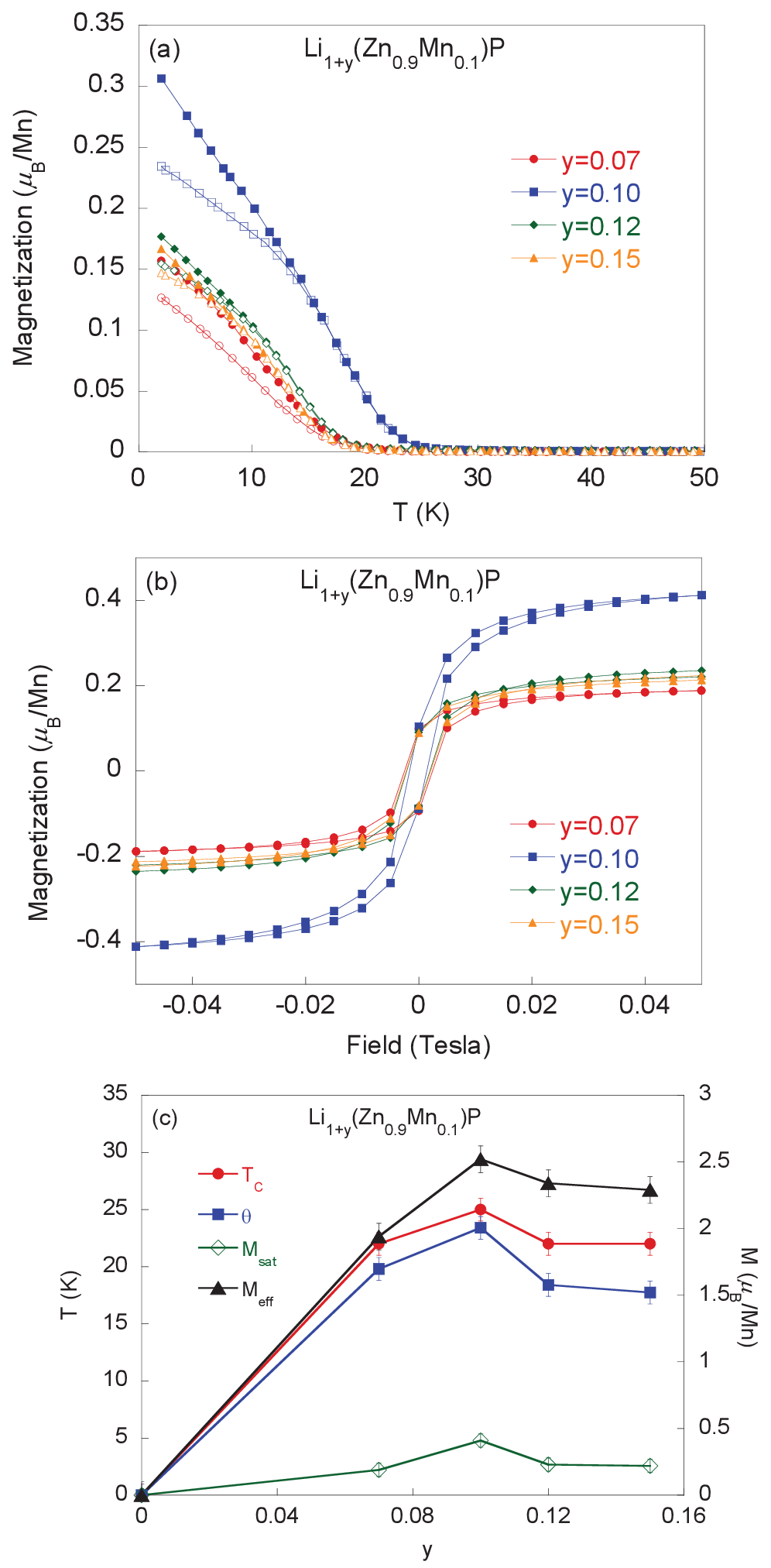}
\caption{(Color online) (a) dc magnetization for
Li$_{1+y}$(Zn$_{0.9}$Mn$_{0.1}$)P ($y=0.07$, 0.10, 0.12, and 0.15)
in both ZFC (open symbols) and FC (solid symbols) modes under
$B_{ext}$ = 50 Oe. (b) Isothermal magnetization curves at 5 K. (c)
$T_{C}$, $\theta$, $M_{sat}$, and $M_{eff}$ versus excess Li$^+$
concentrations y of Li$_{1+y}$(Zn$_{0.9}$Mn$_{0.1}$)P.}
\end{figure}

Li$_{1+y}$(Zn$_{1-x}$Mn$_x$)P has the advantage of decoupled carrier
and spin doping, where the Mn$^{2+}$ substitution for Zn introduces
spins and the Li$^+$ off-stoichiometry provides carriers. It has
been shown that doping Mn$^{2+}$ only in
Li$_{1+y}$(Zn$_{1-x}$Mn$_x$)P, without excess lithium doping, will
result in a paramagnetic ground state \cite{Li(ZnMn)P_JCQ}. Only
with excess Li being doped, can ferromagnetic ordering develop. This
enables us to investigate the influence of overdoped Li on the
ferromagnetic state. In Fig. 6(a) and (b), we show dc magnetization
and isothermal magnetization measured for
Li$_{1+y}$(Zn$_{0.93}$Mn$_{0.07}$)P with a fixed Mn concentration of
$x$ = 0.07 and $y=0.07$, 0.10 and 0.15. We can see that all three
specimens display strong ferromagnetic signals in both $M(T)$ and
$M(H)$ curves. For example, $M(T)$ of
Li$_{1.07}$(Zn$_{0.93}$Mn$_{0.07}$)P shows an strong enhancement
below $T_{C}$ = 26 K ($T_{C}$ is extrapolated from $1/M\sim T$, see
ref. 14), and M(H) has a parallelogram shape with a coercive field
of $\sim$ 20 Oe. Another feature of the
Li$_{1.07}$(Zn$_{0.93}$Mn$_{0.07}$)P specimen is that ZFC and FC
curves display a bifurcation shortly below $T_C$. We fitted the
$M(T)$ curve above $T_C$ according to the Curie-Wess law,
$\chi-\chi_{0}=C/(T-\theta)$, where $C$ is Curie constant and
$\theta$ the Weiss temperature, and derived the effective
paramagnetic moment, $M_{eff}$, from the Curie constant $C$.
$M_{eff}$ is $\sim$ 3$\mu_B$/Mn for $y$ = 0.07. This value is
smaller than 4.5 $\mu_B$/Mn of Li$_{1.04}$(Zn$_{0.97}$Mn$_{0.03}$)P
in ref. 14. $M_{eff}$ $\sim$ 5-6 $\mu_B$/Mn is expected for high
state Mn$^{2+}$ ions. Apparently, the suppression of $M_{eff}$ is
not only arising from the competition of the antiferromagnetic
exchange interaction of N.N. Mn$^{2+}$, but also from the overdoped
carriers, as will be shown in detail in the following.

We plot Curie temperature ($T_C$), Weiss temperature ($\theta$), the
effective moment ($M_{eff}$), and the saturation moment ($M_{sat}$,
the value measured at $T=5$ K and H = 500 Oe) versus the nominal
excess Li concentration in Fig. 6(c). For Li concentration equal to
1 (i.e., y = 0), Li(Zn$_{0.93}$Mn$_{0.07}$)P remains a paramagnetic
ground state. We therefore define T$_C$, $\theta$, M$_eff$ and
M$_sat$ as zero, and plot zero in Fig. 6(c) for comparison. Clearly,
we can see that all four parameters display qualitatively similar Li
concentration dependence. They increase with Li concentration
initially, reach a maximum at $y$ = 0.07, and then start to decrease
when overdoped with Li. To examine if the same trend is valid in a
different Mn concentration, we also investigate these parameters in
Li$_{1+y}$(Zn$_{0.9}$Mn$_{0.1}$)P (with $x$ = 0.10 and $y=0.07$,
0.10, 0.12 and 0.15), and show the results in Fig. 7. The situation
is similar to the case of the $x$ = 0.07 specimens. $T_{C}$,
$\theta$, $M_{sat}$, and $M_{eff}$ first increase from $y=0$ to
$y=0.07$, reach a maximum at $y=0.10$, and then decrease from
$y=0.10$ to $y=0.15$ for Li$_{1+y}$(Zn$_{0.9}$Mn$_{0.1}$)P. We
tabulate all these parameters in Table 1.

It has been theoretically proposed that in diluted magnetic
semiconductors, spins are mediated by hole carriers through RKKY
interaction \cite{ZenerModel}. From NMR measurements
\cite{LiZnMnP_NMR} of Li$_{1.15}$(Zn$_{0.9}$Mn$_{0.1}$)P, it has
been shown that the spin lattice relaxation rate $\frac{1}{T_1}$ of
Li(0) sites, where zero means no Mn at N.N. Zn sites for Li, display
a Korringa behavior, i.e., Fermi surface excitations of a small
number of conduction carriers. The $\frac{1}{T_1}$ of Li(Mn) site
shows a T-independent behavior caused by spin fluctuations, i.e.,
representing local moments. This provides direct and convincing
experimental evidence that Fermi degenerate conduction carriers
mediate the Mn-Mn spin interactions through the p-d exchange
interaction, and that the Mn-Mn spin interaction is long-ranged,
rather than a nearest-neighbor exchange interactions
\cite{LiZnMnP_NMR}. The RKKY exchange interaction can be written as,
J $\sim$ cos(2$k_F$r)/r$^3$, where $k_F$ is the radius of Fermi
surface (assuming the Fermi surface is a spherical shape) and $r$
the distance between two localized moments. The first oscillation
period of the RKKY interaction supports ferromagnetic coupling. In
our case, doping more Li introduces more carrier, and subsequently
modifies the density of states and, therefore, the shape of Fermi
surface. The modified Fermi surface affects the period of
oscillation, as well as the ferromagnetic ordering.

\section{Conclusion}

We conducted $\mu$SR investigation of a bulk form I-II-V DFS
Li$_{1+y}$(Zn$_{1-x}$Mn$_x$)P. Our $\mu$SR results confirm the
development of ferromagnetic ordering below $T_C$ with a nearly
100\% magnetically ordered volume. Despite its much lower carrier
density, Li$_{1+y}$(Zn$_{1-x}$Mn$_x$)P shares a common mechanism for
the ferromagnetic exchange interaction with (Ga,Mn)As, Li(Zn,Mn)As,
1111 type (La,Ba)(Zn,Mn)AsO and 122 type (Ba,K)(Zn,Mn)$_2$As$_2$.
Taking advantage of the decoupled nature of the carrier and spin
doping in Li$_{1+y}$(Zn$_{1-x}$Mn$_x$)P, we investigated, for the
first time, the influence of overdoped carriers on the ferromagnetic
ordered state of a DFS. We found that overdoped Li suppresses both
$T_C$ and magnetic moments. A simple explanation is the modification
of the Fermi surface caused by extra carriers. More detailed
theoretical models are expected to explain this phenomena.

\begin{acknowledgments}
The work at Zhejiang was supported by National Basic Research
Program of China (No. 2014CB921203, 2011CBA00103), NSF of China (No.
11274268); at IOP in Beijing by the NSFC and MOST; at Columbia by
the U.S. NSF PIRE (Partnership for International Research and
Education, Grant No. OISE- 0968226) and Grant No. DMR-1105961; the
JAEA Reimei Project at IOP, Columbia, PSI, and McMaster; and NSERC
and CIFAR at McMaster. F.L. Ning acknowledges helpful discussions
with I. Mazin, I. Zutic.
\end{acknowledgments}


\begin{thebibliography}{References}
\bibitem{InMnAs}Munekata H., Ohno H., Molnar S. von, Segm\"{u}ller
Armin, Chang L. L., and Esaki L., (1989) Phys. Rev. Lett \textbf{63},
1849.

\bibitem{GaMnAs}Ohno H., Shen A., Matsukura F., Oiwa A., Endo A.,
Katsumoto S., and Iye Y., (1996) Appl. Phys. Lett \textbf{69}, 363.

\bibitem{Samarth}Samarth N., (2010) Nat. Mater. \textbf{9}, 955.

\bibitem{Chambers}Chambers S., (2010) Nat. Mater. \textbf{9}, 956.

\bibitem{Dietl1}Dietl T., (2010) Nat. Mater. \textbf{9}, 965.

\bibitem{Dietl2}Dietl T., and Ohno H., (2014) Rev. Mod. Phys. \textbf{86},
187.

\bibitem{Zutic}Zutic I., Fabian J., and Das Sarma S., (2004) Rev.
Mod. Phys. \textbf{76}, 323.

\bibitem{Jungwirth}Jungwirth T., Sinova J., Masek J., Kucera J.,
and MacDonald A.H., (2006) Rev. Mod. Phys. \textbf{78}, 809.

\bibitem{ZenerModel}Dietl T., Ohno H., Matsukura F., Cibert J., and
Ferrand D., (2000) Science \textbf{287},1019.

\bibitem{GaMnAs_190K}Wang M., Campion R. P., Rushforth A. W., Edmonds
K. W., Foxon K. W., and Gallagher B. L., (2008) Appl. Phys. Lett.
\textbf{93}, 132103.

\bibitem{ZhaoJH1}Chen L., Yan S., Xu P. F., Wang W. Z., Deng J. J.,
Qian X., Ji Y., and Zhao J. H., (2009) Appl. Phys. Lett. \textbf{95},
182505.

\bibitem{ZhaoJH2}Chen L., Yang X., Yang F. H., Zhao J. H., Misuraca
J., Xiong P., and Molnar S. V., (2011) Nano Lett. \textbf{11}, 2584.

\bibitem{Li(ZnMn)As_JCQ}Deng Z., Jin C. Q., Liu Q. Q., Wang X. C.,
Zhu J. L., Feng S. M., Chen L. C., Yu R. C., Arguello C., Goko T.,
Ning F. L., Zhang J. S., Wang Y. Y., Aczel A. A., Munsie T.,
Williams T. J., Luke G. M., Kakeshita T., Uchida S., Higemoto W.,
Ito T. U., Gu B., Maekawa S., Morris G. D. and Uemura Y. J., (2011)
Nat. Commun. \textbf{2}, 422.

\bibitem{Li(ZnMn)P_JCQ}Deng Z., Zhao K., Gu B., Han W., Zhu J. L.,
Wang X. C., Li X., Liu Q. Q., Yu R. C., Goko T., Frandsen B., Liu
L., Zhang J. S., Wang Y. Y., Ning F. L., Maekawa S., Uemura Y. J.,
and Jin C. Q., (2013) Phys. Rev. B \textbf{88}, 081203(R).

\bibitem{Li(ZnCr)As_WQ}Wang Q., Man H. Y., Ding C., Gong X., Guo
S. L., Wang H. D., Chen B., and Ning F. L., (2014) J. Appl. Phys.\textbf{
115}, 083917.

\bibitem{(LaBa)(ZnMn)AsO_DC}Ding C., Man H. Y., Qin C., Lu J. C.,
Sun Y. L., Wang Q., Yu B. Q., Feng C. M., Goko T., Arguello C. J.,
Liu L., Frandsen B. A., Uemura Y. J., Wang H. D., Luetkens H., Morenzoni
E., Han W., Jin C. Q., Munsie T., Williams T.J., D\textquoteright{}Ortenzio
R. M., Medina T., Luke G. M., Imai T., and Ning F. L., (2013) Phys.
Rev. B \textbf{88}, 041102(R).

\bibitem{(LaSr)(ZnFe)AsO_LJC}Lu J. C., Man H. Y., Ding C., Wang Q.,
Yu B. Q., Guo S. L., Wang H. D., Chen B., Han W., Jin C. Q., Uemura
Y. J., and Ning F. L., (2013) Europhys. Lett. \textbf{103}, 67011.

\bibitem{(LaSr)(CuMn)SO_YXJ}Yang X. J., Li Y. K., Shen C. Y., Si
B. Q., Sun Y. L., Tao Q., Cao G. H., Xu Z. A., and Zhang F. C., (2013)
Appl. Phys. Lett. \textbf{103}, 022410.

\bibitem{(LaCa)(ZnMn)SbO_JCQ}Han W., Zhao K., Wang X. C., Liu Q.
Q., Ning F. L., Deng Z., Liu Y., Zhu J. L., Ding C., Man H. Y., and
Jin C. Q., (2013) Sci. China-Phys. Mech. Astron. \textbf{56}, 2026.

\bibitem{(BaK)(ZnMn)2As2_JCQ}Zhao K., Deng Z., Wang X. C., Han W.,
Zhu J. L., Li X., Liu Q. Q., Yu R. C., Goko T., Frandsen B., Liu L.,
Ning F. L., Uemura Y. J., Dabkowska H., Luke G. M., Luetkens H., Morenzoni
E., Dunsiger S. R., Senyshyn A., B\"{o}ni P., and Jin C. Q., (2013)
Nat. Commun. \textbf{4}, 1442.

\bibitem{(BaK)(CdMN)2As2_YXJ}Yang X. J., Li Y. K., Zhang P., Luo
Y. K., Chen Q., Feng C. M., Cao C., Dai J. H., Tao Q., Cao G. H.,
and Xu Z. A., (2013) J. Appl. Phys. \textbf{114}, 223905.

\bibitem{32522_MHY}Man H. Y., Qin C., Ding C., Wang Q., Gong X.,
Guo S. L., Wang H. D., Chen B., and Ning F. L., (2014) Europhys. Lett.
\textbf{105}, 67004.

\bibitem{Dunsiger}Dunsiger S. R., Carlo J. P., Goko T., Nieuwenhuys
G., Prokscha T., Suter A., Morenzoni E., Chiba D., Nishitani Y., Tanikawa
T., Matsukura F., Ohno H., Ohe J., Maekawa S., and Uemura Y. J., (2010)
Nat. Mater. \textbf{9}, 299.

\bibitem{LiZnMnP_NMR}Ding C., Qin C., Man H. Y., Imai T., and Ning
F. L., (2013) Phys. Rev. B \textbf{88}, 041108(R).

\bibitem{LiZnPgap}Kuriyama K., Katoh T., and Mineo N., (1991) J.
Cryst. Growth \textbf{108}, 37.

\bibitem{LiZnPgap2}Bacewicz R., and Ciszek T. F., (1988) Appl. Phys.
Lett. \textbf{52}, 1150.

\bibitem{Uemura}Uemura Y. J., Goko T., Gat-Malureanu I. M., Carlo
J. P., Russo P. L., Savici A. T., Aczel A., MacDougall G. J., Rodriguez
J. A., Luke G. M., Dunsiger S. R., McCollam A., Arai J., Pfleiderer
Ch., B\"{o}ni P., Yoshimura K., Baggio-Saitovitch E., Fontes M. B.,
Larrea J., Sushko Y. V., and Sereni J., (2007) Nat. Phys. \textbf{3},
29.




\end{thebibliography}
\end{document}